\documentclass[twoside,twocolumn]{article}
\usepackage{fitee}
\usepackage[colorlinks, breaklinks = true]{hyperref}		
\pdfoutput=1

\addto{\captionsenglish}{%
  
}

\usepackage{xspace}
\usepackage{listings}
\usepackage[normalem]{ulem}
\usepackage{color, xcolor}

\usepackage{graphicx}
\usepackage{subfigure}

\newcommand\etal{\emph{et al.}\xspace}
\newcommand\htSysName{\texttt{hthreads}\xspace}
\newcommand\clSysName{\texttt{MOCL3}\xspace}

\newcommand\cparagraph[1]{\vspace{1mm}\noindent\textbf{#1.}}

\definecolor{mygreen}{rgb}{0,0.6,0}
\definecolor{bluekeywords}{rgb}{0.13, 0.13, 1}
\definecolor{greencomments}{rgb}{0, 0.5, 0}
\definecolor{redstrings}{rgb}{0.9, 0, 0}
\definecolor{graynumbers}{rgb}{0.5, 0.5, 0.5}

\definecolor{Gray}{gray}{0.9}

\begin{document}

\title{Programming Bare-Metal Accelerators with Heterogeneous Threading Models: A Case Study of Matrix-3000}

\author[1]{Jianbin FANG}%
\author[1]{Peng ZHANG}%
\author[$\ddagger$1]{Chun HUANG}%
\author[1]{Tao TANG}
\author[1]{Kai LU}%
\author[1]{Ruibo WANG}%
\author[2]{Zheng WANG}
\affil[1]{School of Computer Science and Technology, National University of Defense Technology, Changsha 410073, China}
\affil[2]{School of Computing, University of Leeds, Leeds LS2 9JT, United Kingdom}

\shortauthor{FANG et al.}	

\authmark{}



\corremailA{\{j.fang, zhangpeng13a, chunhuang, tangtao84, kailu\}@nudt.edu.cn}
\corremailB{ruibo@yeah.net}
\corremailC{z.wang5@leeds.ac.uk}



\abstract{As the hardware industry moves towards using specialized heterogeneous many-cores to avoid the effects of the
power wall, software developers are finding it hard to deal with the complexity of these systems. This article shares
our experience when developing a programming model and its supporting compiler and libraries for Matrix-3000, which is
designed for next-generation exascale supercomputers but has a complex memory hierarchy and processor organization. To
assist its software development, we developed a software stack from scratch that includes a low-level programming
interface and a high-level OpenCL compiler. Our low-level programming model offers native programming support for using
the bare-metal accelerators of Matrix-3000, while the high-level model allows programmers to use the OpenCL programming
standard. We detail our design choices and highlight the lessons learned from developing systems software to enable the
programming of bare-metal accelerators. Our programming models have been deployed to the production environment of an
exascale prototype system.}

\keywords{Heterogeneous computing; Parallel programming models; Performance; Programmability; Compilers; Runtime systems}




\conf{Jianbin FANG and Peng ZHANG has the same contribution.} \support{Project supported by the National Key Research
and Development Program of China (No.~2021YFB0300101), National Natural Science Foundation of China (No.~61972408), and
a UK Royal Society International Collaboration grant.}
\orcid{Jianbin FANG, http://orcid.org/0000-0003-3542-4869}	
\articleType{}

\maketitle

\section{Introduction}
Heterogeneous many-cores are now  commonplace in computer systems~\cite{DBLP:conf/eurographics/OwensLGHKLP05,
DBLP:journals/pieee/OwensHLGSP08}. The combination of a host CPU with a specialized accelerator (e.g., a
general-purpose GPU (GPGPU), FPGA, DSP or NPU) are shown to deliver orders of magnitude performance improvement over
traditional homogeneous CPU setups~\cite{DBLP:conf/isscc/Patterson18}.  The increasing importance of heterogeneous
many-core architectures can be seen from the TOP500~\footnote{https://www.top500.org/lists/top500/} and
Green500~\footnote{https://www.top500.org/lists/green500/} list, where a large number of supercomputers are integrated
with CPUs and accelerators~\cite{DBLP:journals/jzusc/LiaoLYLYLHLFRS18}. Indeed, heterogeneous many-core architecture is widely seen as the building block for the
next-generation supercomputers.

The potential of accelerators can only be unlocked if the software can make good use of the hardware~\cite{DBLP:conf/icpp/FangVS11, DBLP:conf/icppw/ShenFSV12}. Writing and
optimizing code for many-core accelerators is challenging for many application developers. This is because the current
hardware architecture and programming model of accelerators significantly differ from the conventional multi-core
processors~\cite{DBLP:journals/ibmrd/PerezBBL07}. This change has shifted the burden onto programmers and
compilers~\cite{DBLP:conf/pldi/KudlurM08}. In particular, programmers have to manage the hardware heterogeneity and
parallelism and a complex, distributed memory hierarchy~\cite{DBLP:journals/jzusc/ZhaiC18}.

This article shares our experience of designing and implementing a programming model and its supporting compiler and
runtime system for the Matrix-3000 (also coined as MT-3000) heterogeneous many-core accelerator (Section~\ref{sec:mt3k_arch}).
This accelerator is designed
to be a building block
for next-generation exascale prototype supercomputers~\cite{DBLP:journals/ccfthpc/LuWGHLWFTCLLLS22}. While providing
potential high-performance, MT-3000 has a complex memory hierarchy and processor organization. Furthermore, as no
operating system runs on the accelerator, it is highly challenging to debug and manage parallel threads running on the
accelerators.  In a nutshell, the architecture design of MT-3000 poses a range of challenges to low-level systems software design.

To support software development for MT-3000, we have developed a full-stack system software (Section~\ref{sec:stack})
from the scratch, where we focus on two programming modules: the \htSysName low-level programming interface
(Section~\ref{sec:hthreads}) and the \texttt{MOCL3} OpenCL compiler (Section~\ref{sec:mocl3}). Specifically, we develop
a low-overhead and high-availability heterogeneous programming interface (\texttt{hthreads}). At the core of \htSysName
is a heterogeneous threading model introduced for the bare-metal MT-3000 accelerator.
The \htSysName interface consists of the general-purpose zone side APIs and the acceleration zone side APIs. On the one
hand, \htSysName exposes as many performance-related architecture features as we can, aiming to fully tap its compute
potentials.
On the other hand, we introduce the threading model to hide the metal-related uses such as the 
native DMA usage.
At a higher level, we provide the implementation of OpenCL standard parallel programming interface (\texttt{MOCL3}) for the MT-3000 architecture.
It follows the programming specification of OpenCL version 1.2.
While ensuring to explore the computing potential of MT-3000, \clSysName can be effectively compatible with
OpenCL legacy code and significantly improve programmability.
With these two programming interfaces, we aim to achieve a balance between performance, programmability, and portability for our
home-grown bare-metal accelerator.

Both \texttt{hthreads} and \texttt{MOCL3} have been deployed to the production environment of an exascale prototype
system with the MT-3000 executing
\texttt{hthreads} and \texttt{MOCL3} optimized code at any time.
We showcase the performance of the developed system on micro-benchmarks and matrix multiplications (Section~\ref{sec:results}).
We hope that the experience presented in
this article can provide new insights into the development of future high-performance accelerators and their
programming systems.

\section{The Matrix-3000 Architecture} \label{sec:mt3k_arch}
This section describes the hardware architecture design of the MT-3000 accelerator targeted in this article.
\subsection{Heterogeneous Multi-zones\label{sec:mz}}
\begin{figure}[!t]
\centering
\includegraphics[width=0.98\linewidth]{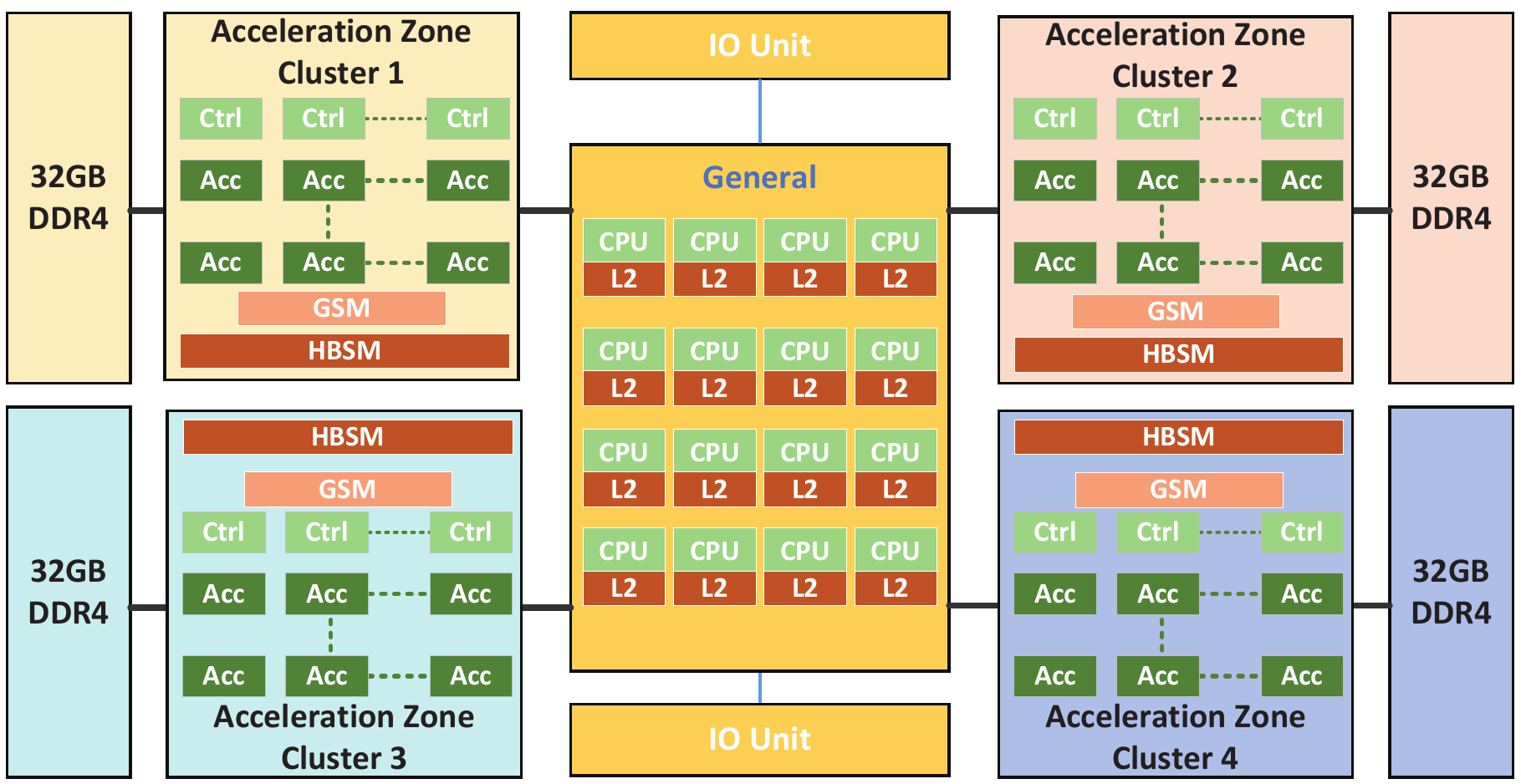}
\caption{Overview of the Matrix-3000 architecture.}
\label{fig1:MT-3000}
\end{figure}
As depicted in Figure~\ref{fig1:MT-3000},
MT-3000 implements a multi-zone microarchitecture
with 16 CPU cores, 96 control cores, and 1536 accelerator cores.
The CPU cores
form a general-purpose zone (GP zone), while the control
cores and the accelerator cores
form an acceleration zone (ACC zone).
The acceleration zone is then equally divided
into four autonomous acceleration clusters.
Each cluster has 24 control cores, 384 accelerator cores,
and on-chip global shared memory (GSM), high-bandwidth shared memory (HBSM), and off-chip DDR memory.
The GP zone CPU cores run at 2.0 GHz while the ACC zone cores operate at 1.2 GHz and can deliver a total of 11.6
Tflops double-precision performance with a power efficiency of 45.4 Gflops per watt.
The four ACC zone clusters can run independently
of each other.


The GP zone runs an operating system. Different from the GP zone, the acceleration cluster is a bare-metal device,
with no support of operating system,
and all hardware resources need to be managed by user programs.
The CPU cores in the general-purpose zone are capable of managing the overall task execution,
running the operating system, and processing general-purpose tasks,
while the acceleration zone is designed for computation-intensive tasks.
The CPU and the accelerator have different accessing scopes of the memory hierarchy, which are detailed in the following subsection.

%


\subsection{Hybrid Memory Hierarchy}

MT-3000 is featured with a hybrid memory hierarchy.
Processing cores in the general-purpose and the acceleration zones have
different accessing
scopes of the memory system.

\cparagraph{General-purpose zone} Each of the 16 CPUs in the general-purpose zone has its own L1 and L2 cache.
And the CPU cores are connected through a mesh-based on-chip network (NoC) that achieves cache coherency.
The L2 cache size is 512KB, which is organized in a 16-way set associative with a 64-byte cacheline.
The L1 cache adopts an inclusive policy.
Therefore,  when an L2 data element is requested by other CPUs, the L2 cache controller will have to
retrieve the newest values from its corresponding L1 cache if the data is dirty.
The CPUs support optional prefetch, i.e.,
upon a cache miss, 0, 2, 4, or 8 cachelines would be prefetched into L2.

\cparagraph{Acceleration zone}
As shown in Figure~\ref{fig1:MT-3000},
each cluster in the acceleration zone
has its own GSM, HBSM and DDR memory. The GSM and HBSM is private to the cluster, which are shared by all the control and
accelerator cores within the same acceleration cluster.
To reduce memory conflicts, both HBSM and GSM are organized into multiple banks.
Combined with direct memory access, such a multi-banked organization provides flexible support for the run-time data
management.

The CPU core can access the entire HBSM/GSM and the DDR space located in different acceleration clusters.
By contrast, accelerator cores can only access the GSM, HBSM and DDR within their acceleration cluster.
Data transfers across different acceleration clusters have to be managed by CPUs.

%
%
%

\begin{figure}[!t]
\centering
\includegraphics[width=0.6\linewidth]{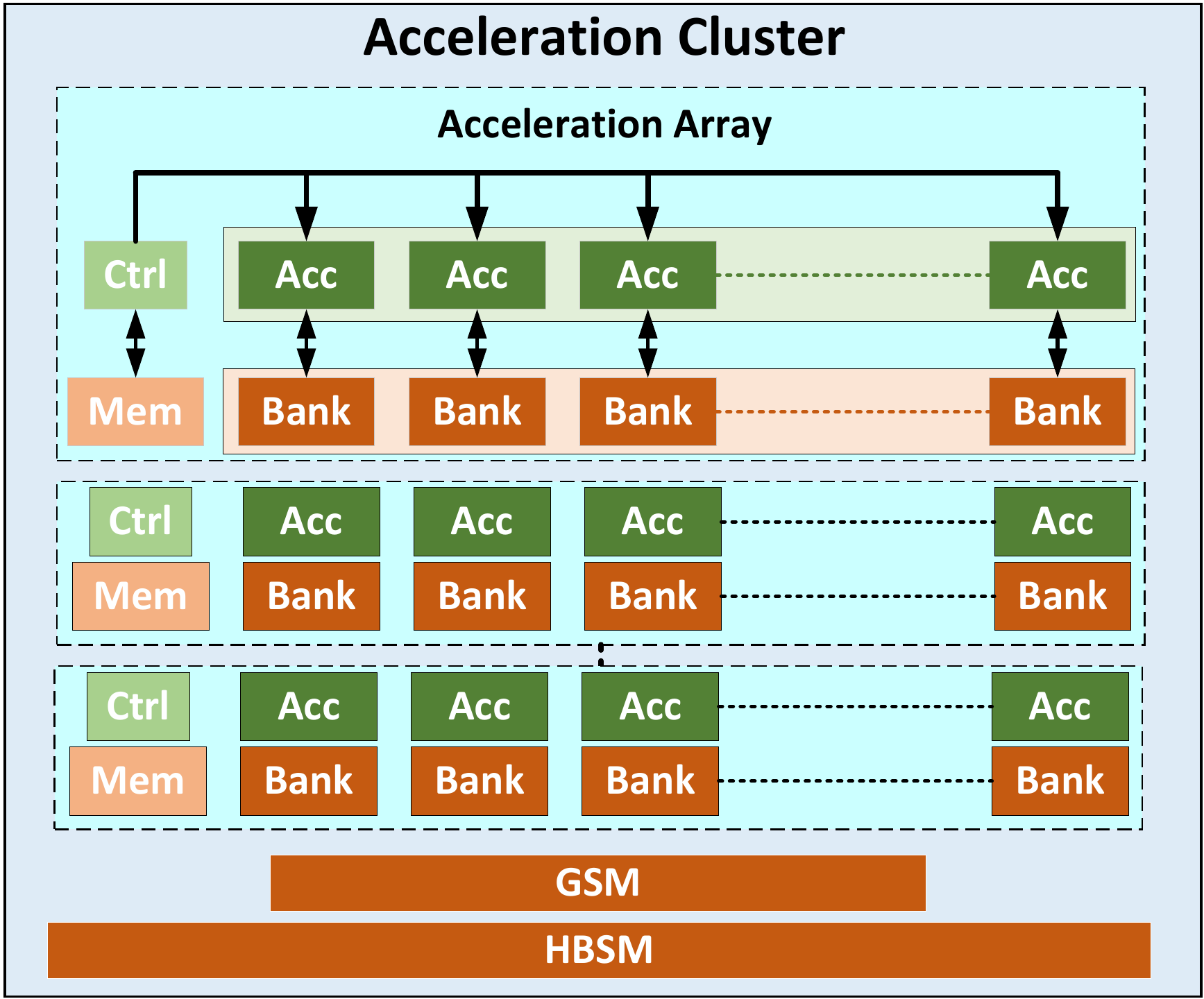}
\caption{The organization of an acceleration cluster.}
\label{fig2:VLIW-SIMD}
\end{figure}

\subsection{Acceleration Array Organization}
\subsubsection{Combined VLIW and Acceleration Array}
Figure~\ref{fig2:VLIW-SIMD} show that each MT-3000 acceleration cluster has 24 acceleration array, each of which further has
one control core and 16 accelerator cores.
In the acceleration array,
the 16 acceleration cores are driven by one single instruction stream and work in a lock-step manner.
The single instruction
stream is handled by the control core. The acceleration array
is the main source for the data level parallelism (DLP),
which is
commonly seen in HPC workloads.

Further, each accelerator core uses a Very Long Instruction Word (VLIW) organization, having
3 multiply and accumulate (MAC) units, one integer execution unit (IEU), and 2 load/store units.
At most 6 instructions can be packed and issued simultaneously to an accelerator core within a cycle.
Each MAC unit supports both fixed and floating-point multiply and accumulate.
For the floating point operations,
the MAC unit supports half, single and double precision floating point operations.
The IEU can support both bitwise and integer operations.
By combining VLIW and acceleration array organization,
we can exploit both data level parallelism and instruction level parallelism.

\subsubsection{On-chip Memory Design}
MT-3000 uses a high-bandwidth on-chip memory design, i.e., scalar memory (SM) and array memory (AM), which is shown in Figure~\ref{fig2:VLIW-SIMD}.
The SM is private to a control core and the AM is shared by the 16 accelerator cores.
AM supports at most two load/stores to each accelerator core.
The data types of AM load/stores include half-word (32-bit), word (64-bit), and double word (128-bit).
Thus, AM can provide at most 512 bytes ($16\times 2 \times 128$ bits) to 16 accelerator cores.
Each SM buffer is of 64KB, which is private to a control core, and each AM buffer is of 768KB, which is private to the 16 accelerator cores.
Note that the AM buffer and the SM buffer are located at the same level.


To summarize, MT-3000 is a bare-metal heterogeneous processor with a complex core and memory organization. Such
hardware design provides the potential for high-performance by giving a large degree of flexibility for the systems
software to optimize the data placement, thread communications and parallel computation. However, this architectural
design requires having an effective programming model to lower the programming difficulties. Our work will focus on
addressing the programming issues of MT-3000 and accelerators of its kind.

\section{The MT-3000 Programming Stack} \label{sec:stack}


\begin{figure}[!t]
\centering
\includegraphics[width=0.725\linewidth]{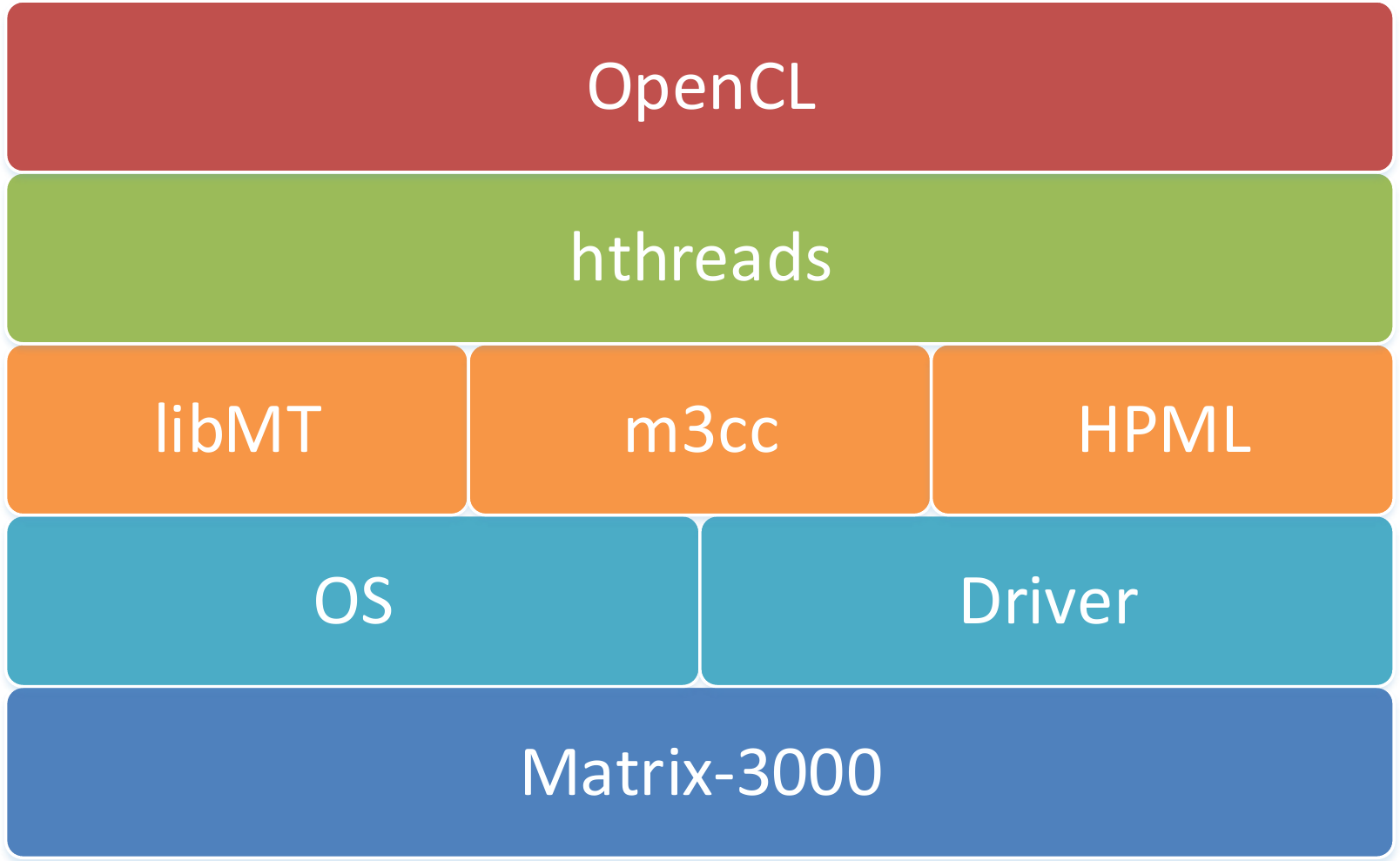}
\caption{The MT-3000 programming stack.}
\label{fig4:toolchain}
\end{figure}

Figure~\ref{fig4:toolchain} gives an overview of our four-layer software stack designed for MT-3000. A standard Linux
operating system (OS) runs on the CPU. The  OS manages the interactions between CPUs and the bare-metal accelerator
clusters in the acceleration zone through a device driver.
The MT-3000 compiler, \texttt{m3cc}, translates C code into executable binaries to run on the MT-3000. As part of the
compilation toolchain, \texttt{libMT} provides a low-level interface for the runtime to manage accelerators, and
\texttt{HPML} is a high-performance math library specifically optimized for MT-3000.

At the core of this work is the two programming interfaces that we have developed for MT-3000. The \texttt{hthreads}
programming model provides a heterogeneous threading model for MT-3000. Building upon hthreads, \texttt{MOCL3}
implements the OpenCL heterogeneous programming standard of version 1.2 for {MT-3000}. In a nutshell, \texttt{hthreads}
and \texttt{MOCL3} are built on top of the lower-level components of the software stack of MT-3000. 

\subsection{The m3cc Compiler}
As a key component of the MT-3000 software stack, we develop a low-level, in-house compiler (\texttt{m3cc}), which translates C99
compatible programs into executable MT-3000 binaries. \texttt{m3cc} is a cross-compiler that runs on the general-purpose CPU of MT-3000 to
generate binaries for the accelerator. In addition to standard C, m3cc supports vectorization intrinsics and embedded assembly codes.
\texttt{m3cc} is integrated with the assembler (\texttt{m3cc-as}) and the linker (\texttt{m3cc-ld}), which form a complete compiling tool
chain for MT-3000.




\subsection{Low-Level Software Interface}
The \texttt{libMT} library acts as an low-level software interface for managing the interactions between the CPUs and the accelerators.
This library first provides the functions of managing the shared buffers and data transfers between CPUs and
accelerators. Given that the accelerators can only use the physical addresses and the CPUs use the virtual addresses,
\texttt{libMT}  has to perform the address translation between them based on the lower-level driver module. Then,
\texttt{libMT}  loads a program image and its kernel argument data onto a predefined location, and fires the
accelerators for kernel execution. As there are caches on the CPU side, \texttt{libMT} also provides interfaces to
maintain the data consistency between the CPUs and the accelerators
by invalidating the data cachelines when needed.

\subsection{High-Performance Math Library}
\texttt{HPML} is a bundle of mathematical libraries, including libm, BLAS, SparseBLAS, FFT, and many others. The math
libraries are highly optimized by experts for the accelerators of MT-3000, aiming to tap the compute potentials of
MT-3000. They are typically hardcoded kernels in assemblies. On the host side, they are implemented in \texttt{libMT} or \htSysName
to manage the interactions between the GP zone and the ACC zone.

\section{The \texttt{hthreads} Programming Interface} \label{sec:hthreads}
\subsection{Design Overview}
To avoid users' directly dealing with the underlying hardware,
and improve the programmability,
we present a low-overhead and high-availability \textbf{h}eterogeneous \textbf{threads} (\texttt{hthreads}) programming interface.

\begin{figure}[!t]
\centering
\includegraphics[width=0.7\linewidth]{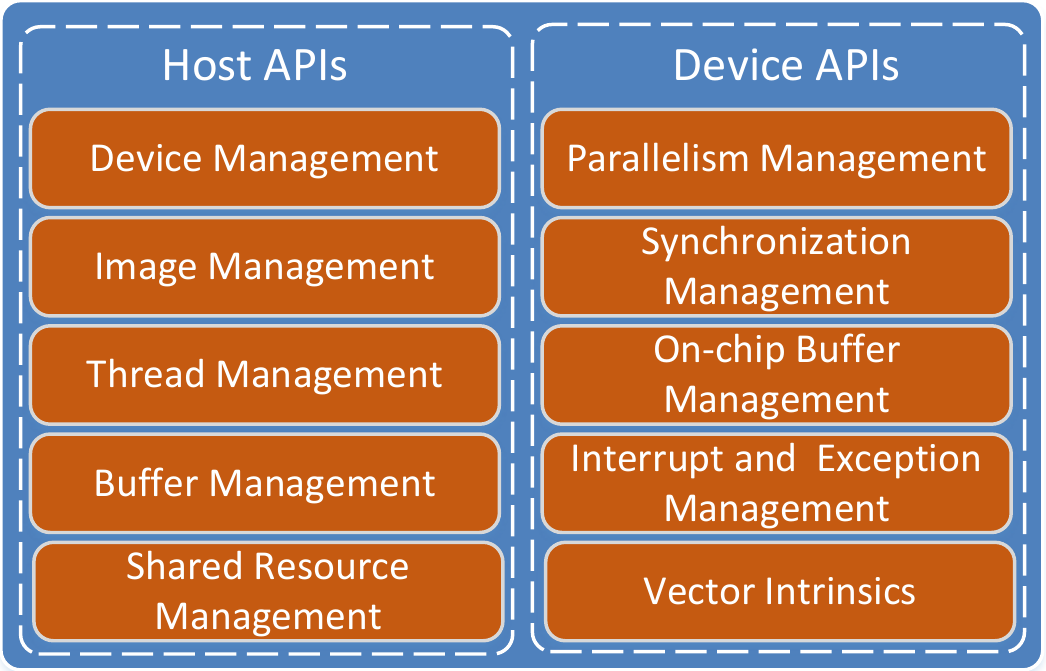}
\caption{Overview of the hthreads interfaces.}
\label{fig:ht_ov}
\end{figure}

Figure~\ref{fig:ht_ov} shows that \htSysName consists of the general-purpose zone side APIs (Host APIs) and the acceleration zone side APIs (Device APIs).
In general, \htSysName takes the GP side as \emph{host}, and takes each acceleration cluster as a \emph{compute device}.
On the host side, we provide APIs to manage devices, program images, threads, and shared resources.
At the core of \htSysName is the introduction of the \emph{threading} concept.
Thus, programmers can use the logical threading instance, rather than the physical cores related concepts.
On the device side, we provide APIs to manage thread parallelism, synchronization, data movements between the on-chip and off-chip buffers, and
the orchestration of the 16 acceleration cores with vector data types and intrinsics.

\cparagraph{An example code in hthreads}
Figure~\ref{fig:code_example} shows a code example for vector addition ($\mathbf{C}[\ ]=\mathbf{A}[\ ]+\mathbf{B}[\ ]$) in \texttt{hthreads}.
We see that hthreads uses \texttt{hthread\_dev\_open} and \texttt{hthread\_dev\_close} to
switch on and off the device, and makes the necessary initialization work.
We use \texttt{hthread\_dat\_load} to load the compiled kernel image into the predefined location (Line~3).
Then it allocates buffers for a specific device with \texttt{hthread\_malloc} (Lines~4--6), and they are freed before the program exits (Line~10).
Programmers have to explicitly specify the buffer size and properties (\texttt{HT\_MEM\_RO}, \texttt{HT\_MEM\_WO} or \texttt{HT\_MEM\_RW}).
After preparing the arguments, we create a \textit{thread group} and launch kernel execution with \texttt{hthread\_group\_create} (Line~7).
When coding kernels, we use a quanlifier \texttt{\_\_global\_\_}.
To manage threads on the device side, we use \texttt{get\_thread\_id} to identify a thread, which is pinned to an acceleration array (Figure~\ref{fig2:VLIW-SIMD}).
We then use \texttt{vector\_\{malloc|free\}} and \texttt{vector\_\{load|store\}} APIs for the AM management and data movements.
These interfaces encapsulate the hardware details of the DMA units with abstractions.

\lstset{}
\begin{figure}
        \noindent\mbox{\parbox{\columnwidth}{%
\begin{lstlisting}[label=subfig:source_in]
// main entry for vector addition
hthread_dev_open(dev_id);
hthread_dat_load(dev_id, "vadd_kernel.dat"); 
long *A = hthread_malloc(dev_id, size, HT_MEM_RO);
long *B = hthread_malloc(dev_id, size, HT_MEM_RO);
long *C = hthread_malloc(dev_id, size, HT_MEM_RW); 
int tg_id = hthread_group_create(dev_id, num_threads, "add_vector", 2, 3, args);
hthread_group_wait(tg_id);
hthread_group_destroy(tg_id);
hthread_free(A); // the same for B and C
hthread_dev_close(dev_id);
/* borderline between host and device codes */
__global__ void add_vector(unsigned long length, long *A, long *B, long *C){
    int tid = get_thread_id();
    long i = 0, b = 0;
    long step_size = 16 * 1024;
    long num_steps = length/step_size;
    unsigned long offset = tid * length;
    lvector long * src1 = vector_malloc(step_size * sizeof(long));
    lvector long * src2 = vector_malloc(step_size * sizeof(long));
    lvector long * dst = vector_malloc(step_size * sizeof(long));
    for(b = 0; b < num_steps; b++){
        vector_load(&A[offset], src1, step_size * sizeof(long));
        vector_load(&B[offset], src2, step_size * sizeof(long));
        for (i = 0; i < step_size / 16; i ++)
            dst[i] = src1[i] + src2[i];
        vector_store(dst, &C[offset], step_size * sizeof(long));
        offset = offset + step_size;
    }
    vector_free(src1);
    vector_free(src2);
    vector_free(dst);
}

\end{lstlisting}%
        }}
    \caption{An example in hthreads for vector addition.}%
    \label{fig:code_example}%
\end{figure}
\subsection{Implementation}
\begin{figure}[!t]
\centering
\includegraphics[width=0.8\linewidth]{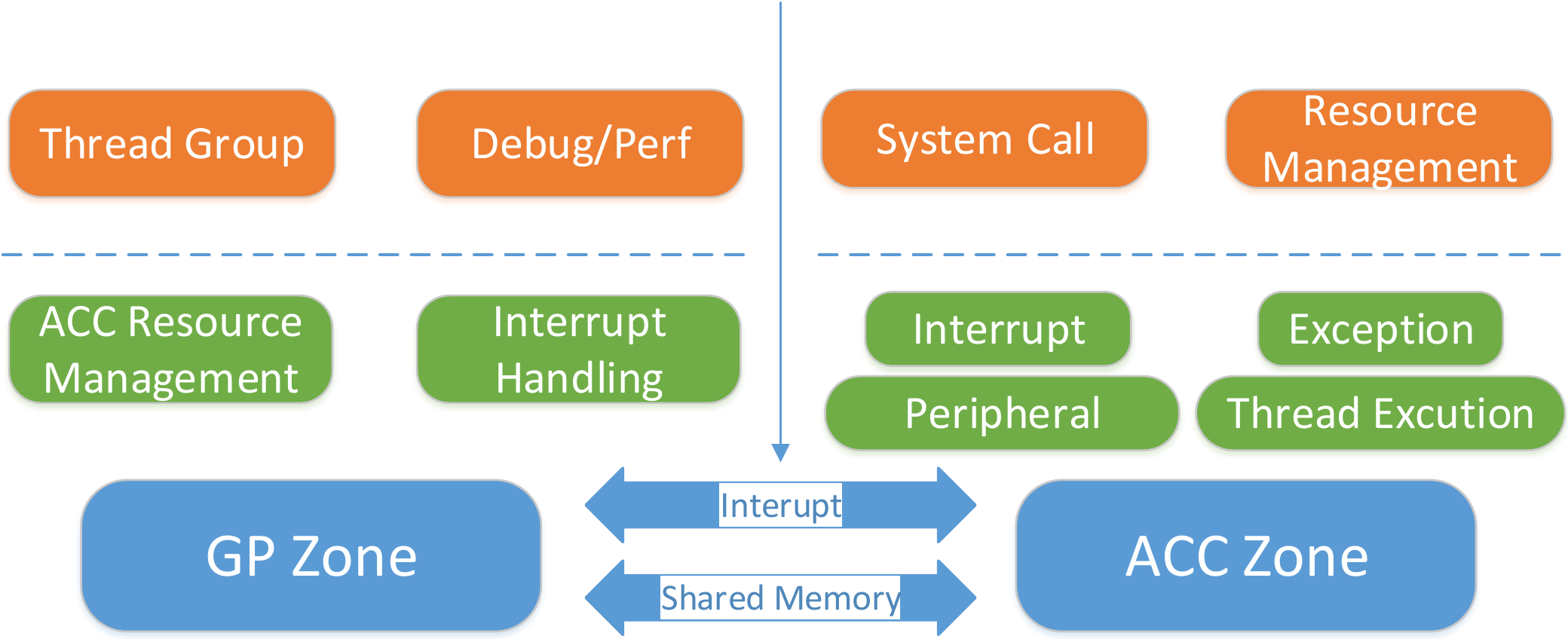}
\caption{The hthreads implementation modules.}
\label{fig:hthreads}
\end{figure}

Figure~\ref{fig:hthreads} shows the conceptual framework of \htSysName.
The module above the dotted line is the interface provided to users,
and the internal implementation function modules are below the dotted line.
The general-purpose zone (GP Zone) and the acceleration zone (ACC Zone) communicate through interrupt and shared memory.



\cparagraph{Memory layout}
As the acceleration clusters have no support of operating system, \htSysName has
to prepare the execution contexts for kernels.
The kernel execution can be treated as an independent process, which has its own process space.
Before kernel execution, \htSysName will load the code segments and data segments into predefined memory locations,
setup the stack register, the entry point, and pass parameters to kernel functions according to the calling conversion (Line~3 of Figure~\ref{fig:code_example}).
Thus we need to set up the memory layout for them.
For the use convenience, we also have to map HBSM, GSM, AM and SM into the process space.

Due to the complex memory hierarchy, the mapping space of memory layout is particularly large.
We can choose to place the scalar stack on DDR, GSM, HBSM or SM.
The code and data segments can also be mapped onto DDR, GSM or HBSM.
When we put the scalar stack on SM,
the instruction latency will be short.
But the SM has limited capacity and it could be too small to hold large codes.
Alternatively, when we put the scalar stack on DDR, the latency will be large.
Therefore, we have to make a tradeoff between performance and buffer capacity.

For the accelerator cores of the ACC zone, we place the heap and stack on the private vector memory of the acceleration array.
Note that the stack space grows downwards and the heap space grows upwards.
As we do not support thread switching, a thread will occupy an entire acceleration array in an exclusive manner until it exits.

\cparagraph{Communication between GP and ACC zones}
Since \texttt{libmt}
is the library of the GP zone, there is no way to communicate between GP zone and ACC zone.
We need implement the communication driver in \htSysName.
We choose not to use an daemon process to act as a management process
in the ACC zone.
Instead,
the process/thread running on each control core has to manage its hardware resource independently.
This means that the GP zone needs to communicate with all the used control cores.

We build the communication driver based on interrupt system to support real-time interactions between the GP zone and the ACC zone.
On the ACC side, we implement a complete interrupt and exception system to handle the relevant events.
This system supports the sending and receiving of interrupts to/from other control cores or the GP zone.
On the GP side, \htSysName supports to send and receive interrupts to/from all the control cores.

There are two ways to implement the interrupt mechanism:
(1) to simulate it by busy waiting, or (2) to implement it based on hardware features of MT-3000.
The former does not require hardware support but consumes computing resources,
whereas the latter does not consume computing resources but requires hardware support.
We implement both approaches, and compare their performance.

\begin{figure}[t]
\centering
\subfigure[Direct model.]{\includegraphics[width=0.8\linewidth]{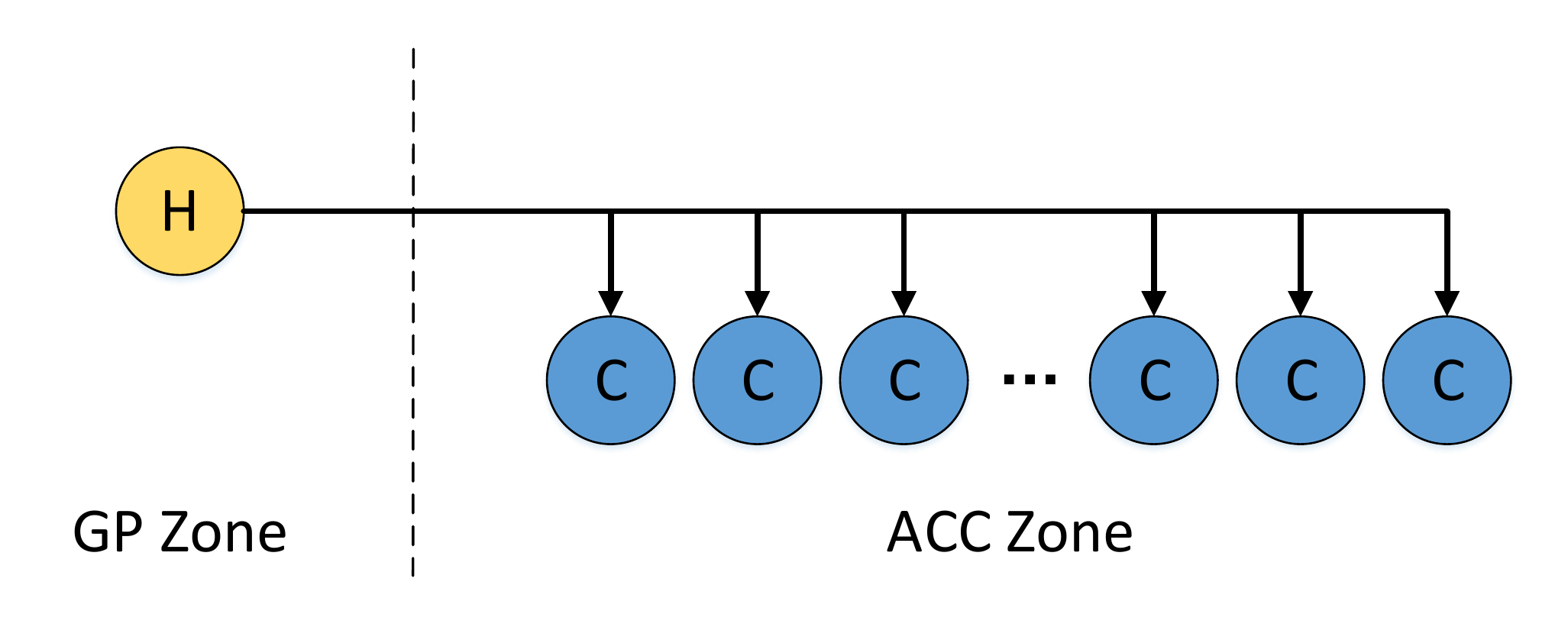} \label{fig:hthreads_1m}}
\subfigure[Proxy model.]{\includegraphics[width=0.8\linewidth]{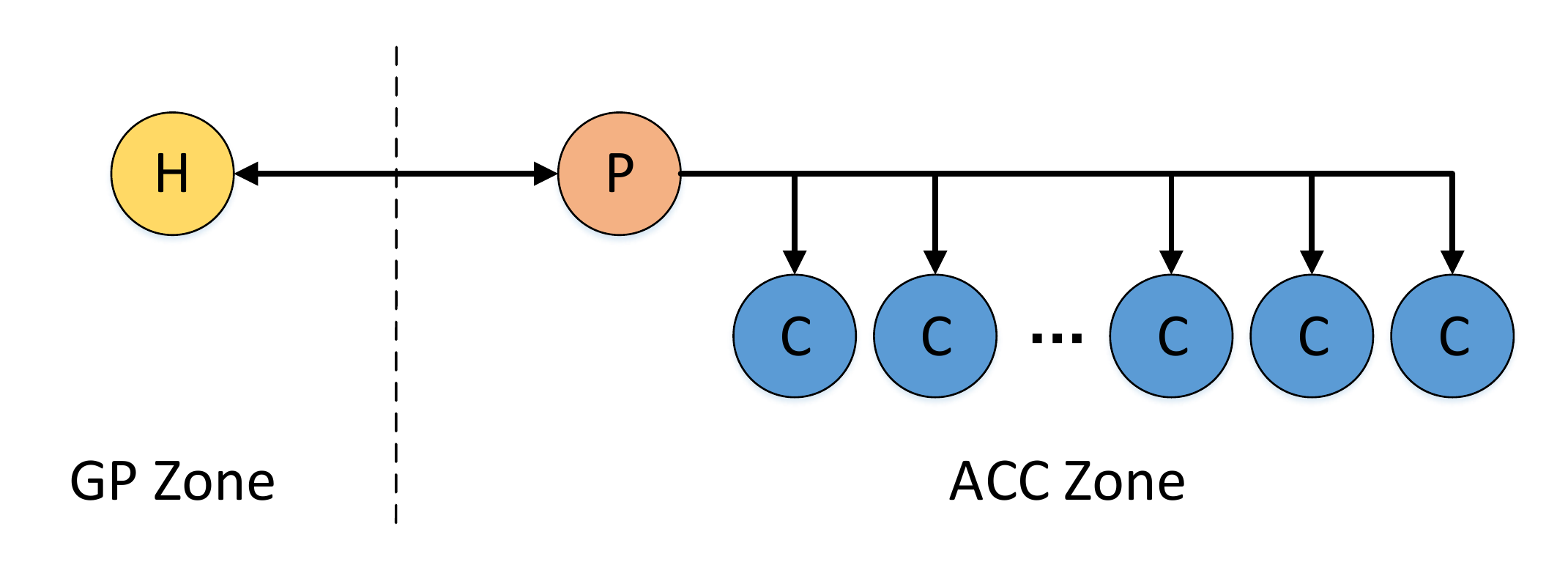} \label{fig:hthreads_2m}}
\caption{The hthreads communication model: (a) The host thread in GP zone communicates with all acc cores directly, (b) The host thread in GP zone only communicates with the proxy core in ACC zone, which then communicates with othe acc cores.}
\end{figure}

To reduce the synchronization cost between the GP zone and the ACC zone,
we design an optimized synchronization model.
Figure~\ref{fig:hthreads_1m} is the direct synchronization model, and 
Figure~\ref{fig:hthreads_1m} is the proxy synchronization model.
In the direct synchronization model, the host thread in GP zone needs to synchronize with
all the ACC cores.
In the proxy synchronization model, the proxy core synchronizes with other cores, then it synchronizes with the host thread in GP zone.
The synchronization overhead between the ACC cores is smaller than the synchronization overhead between the GP zone
and the ACC zone.
So 
it performs better than direct synchronization model, and consumes fewer GP zone resources.

\begin{figure}[!t]
\centering
\subfigure[Naive ACC runtime.]{\includegraphics[width=0.46\linewidth]{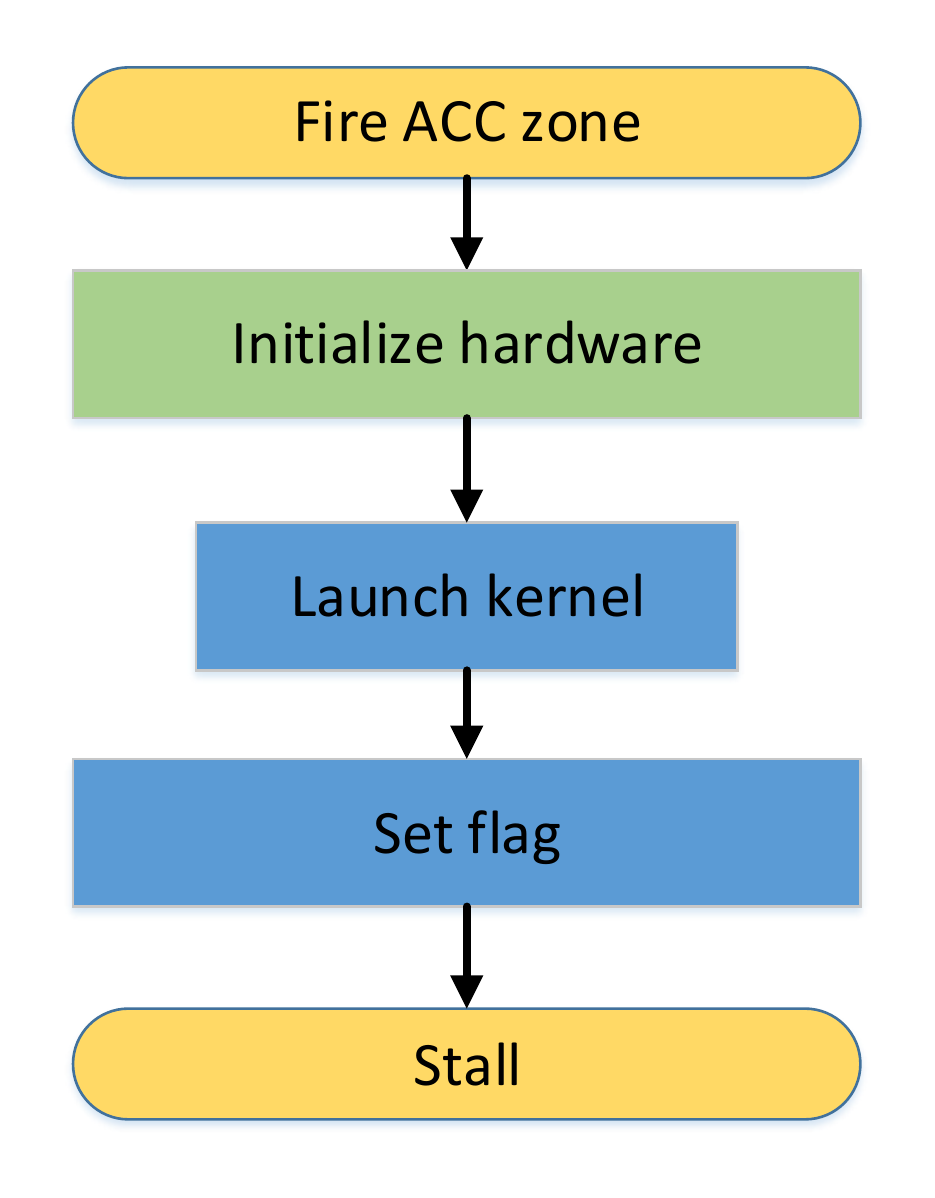} \label{fig:hthreads_d1}}
\subfigure[OS-like ACC runtime.]{\includegraphics[width=0.46\linewidth]{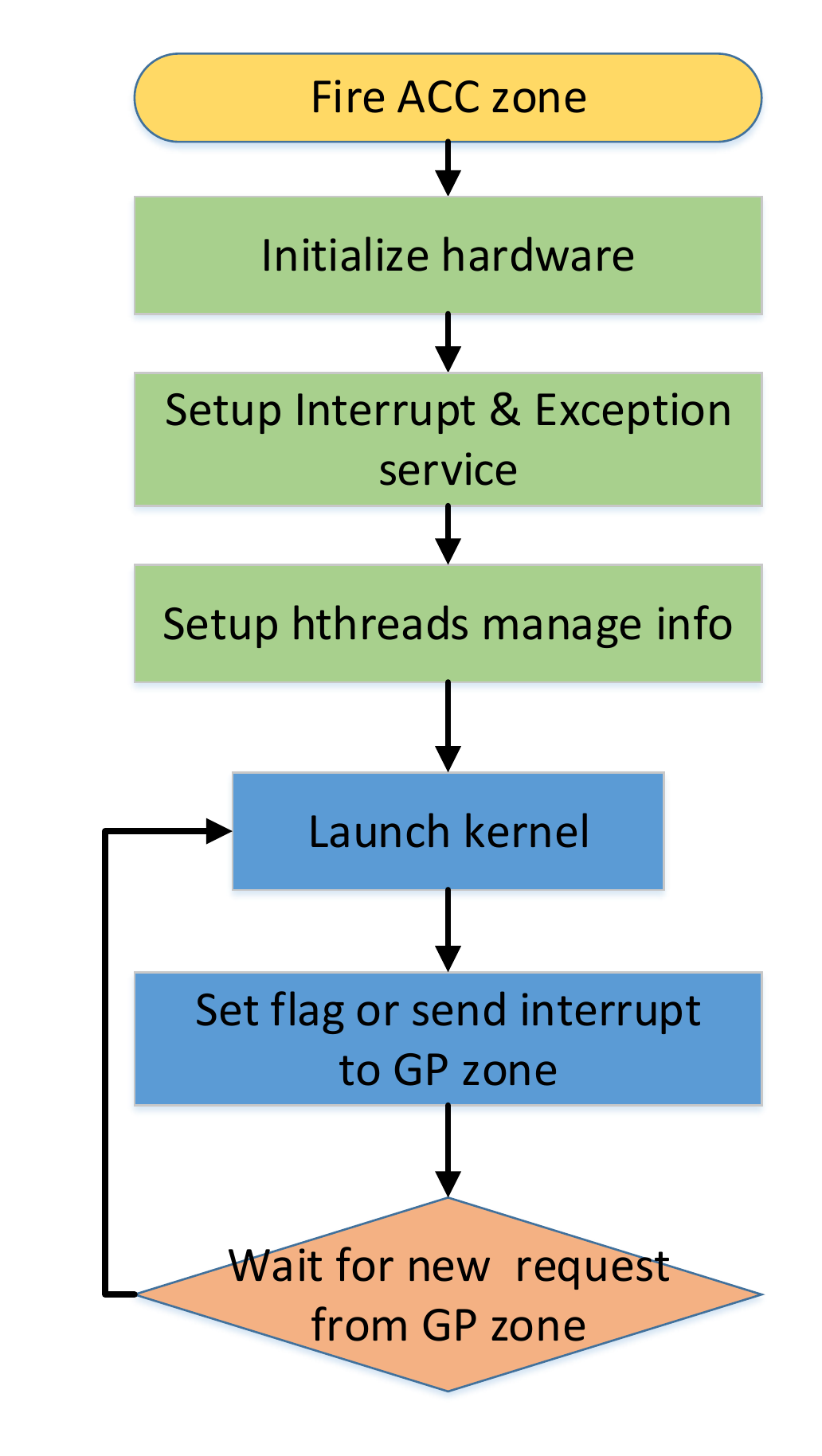} \label{fig:hthreads_d2}}
\caption{The ACC zone runtime: (a) The naive runtime will stall after the user task completed, (b) The OS-like runtime will still run on the ACC zone after the user task completed, and waits for new tasks.}
\end{figure}

\cparagraph{ACC zone runtime}
To minimize the kernel launching overhead, we implement an OS-like runtime for the ACC zone.
Figure~\ref{fig:hthreads_d1} shows the naive ACC zone runtime, which supports kernel execution, but will stall
after kernel completed.
With this naive runtime, users have to fire the ACC zone for each kernel executing, and suffer from a large overhead.
To avoid switching on/off the ACC zone again and again,
we implement an OS-like runtime for the ACC zone, shown in Figure~\ref{fig:hthreads_d2}.
We see that it not only supports kernel execution, 
but setups the interrup and exception services which can report kernel runtime errors.
In addition, the OS-like runtime will still run on the ACC zone after kernel completed, 
and wait for new request from the GP zone.
A new request can be a new kernel execution, or other operations such as synchronization.
A detailed evaluation of the kernel launching overhead will be shown in Section~\ref{sec:overhead}.
We will see that the OS-like runtime can yield a better performance than the native one.

\cparagraph{Vector extension}
To orchastrate the use of the 16 acceleration cores of an acceleration array, we extend standard C to support vector data types and vector operations.
The use of vector data types are demonstrated in Lines~19--21 of Figure~\ref{fig:code_example}, where 
the quanlifier \texttt{lvector} is followed by the relevant conventional data types.

These vector extensions are designed to give the programmer the ability to explicitly control the 16 accelerating cores working in a lock-step manner.
They can also be used by the \texttt{m3cc} compiler to exploit the potential of the acceleration array in the acceleration zones of MT-3000.

\cparagraph{Debugging supports}
As MT-3000 is a new heterogeneous many-core accelerator,
and its ACC zone is a bare-metal device,
programming the ACC zone is error-prone and time-consuming.
And the debugging process requires a deep understanding of the hardware architecture.
To help users with the debugging process, we provide a \texttt{printf} function on the ACC side.
When there occurs an exception from device kernels, \htSysName will print the kernel call stack on the GP side.
And we also implement a \emph{gdb-like} tool (i.e., \texttt{et\_ctl}) to help users debug their codes.
It supports common debugging functions, such as setting breakpoints,
showing the contents of registers or memory, and step execution. 
We are currently enriching the tool to capture more information. 



\section{MOCL3} \label{sec:mocl3}

\texttt{MOCL3} is the implementation of OpenCL standard parallel programming interface for the MT-3000 architecture.
It follows the programming specification of OpenCL (version 1.2).
In general, the implementation of the OpenCL programming model for MT-3000 includes two parts:
the kernel compiler and the runtime system.
The kernel compiler compiles OpenCL kernels into MT-3000 binaries,
and the runtime system implements the programming interfaces defined by the OpenCL specification.
\texttt{MOCL3} is an upgrade for MT-3000 of the OpenCL programming system (\texttt{MOCL})
developed originally for Matrix-2000~\cite{DBLP:conf/cf/ZhangTFHYW18, DBLP:journals/ijpp/JaaskelainenLSR15}.

\subsection{The OpenCL Kernel Compiler}

From the programmer's point of view, an OpenCL program includes two parts:
host side code and device side code (i.e., \emph{kernels}). When compiling a kernel,
we need to compile the OpenCL C code into MT-3000 binaries.
The OpenCL kernel is written according to OpenCL C (based on C99) specification,
but it also has syntax extensions and constraints.

\begin{figure}[!t]
\centering
\includegraphics[width=0.7\linewidth]{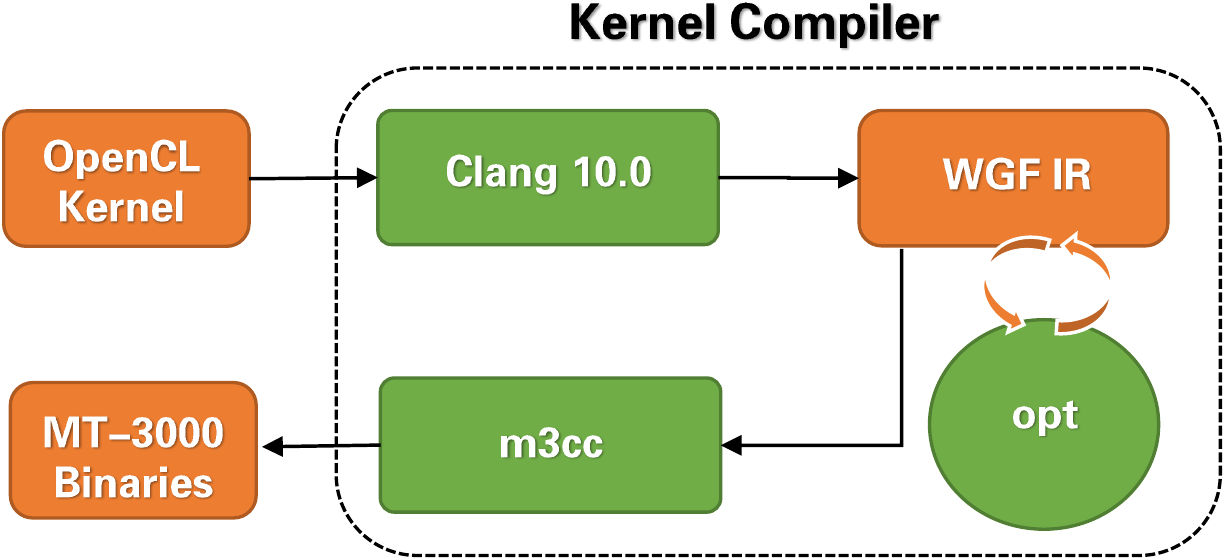}
\caption{The MOCL3 kernel compiler.}
\label{fig:ocl:compiler}
\end{figure}

Figure~\ref{fig:ocl:compiler} shows that our kernel compiler for MT-3000 is implemented in three steps.
We first convert OpenCL kernels into workgroup functions with a loop (\emph{WGF}), represented in LLVM IR.
According to the index space defined by the OpenCL program, the translated program is the task to be performed by a single workgroup function.
Different workgroup functions share the same code, but access different data elements through the index space.
Note that, we perform optimizations on the WGF IR codes with a customized optimizer. 
Due to the lack of LLVM backend for MT-3000, we use the \texttt{llvm-cbe} tool~\footnote{https://github.com/JuliaComputing/llvm-cbe/} to translate the WGF IRs into C codes. 
Third, we compile the workgroup functions into MT-3000 binary representations with the m3cc compiler.

\cparagraph{Handling local variables}
The local variables of OpenCL kernels are shared among all the work-items of the same work-group. 
We convert such local variables to an additional work-group function argument with a fixed allocation size.
During runtime, the local variables are mapped to a predefined address of the on-chip buffers, i.e., either SM or AM (Figure~\ref{fig2:VLIW-SIMD}). 
Given that programmers have to manually move data between on-chip buffers and off-chip DDR memories, 
we implement the relevant builtins (\texttt{async\_work\_group\_copy}) to assist the data movements between local memories and global memories.

\cparagraph{Atomics implementations}
The OpenCL C provides atomic operations to locations in \texttt{\_\_global} or \texttt{\_\_local} memory.
In \texttt{MOCL3}, a work-group is translated  into a 
work-item loop by our kernel compiler, which is scheduled to a hardware thread. 
The work-items within a work-group are executed one by one and in a sequential fashion. 
In terms of memory access, the work-items of this work-group will access local variables sequentially. 
Therefore, the atomic operations on local memories can be replaced by equivalent functional operations without synchronization.
For the global memory case, we reply on the hardware locks of MT-3000 to implement the atomic functions. 
Again, \texttt{hthreads} provides relevant APIs to manage the shared resource such as \emph{locks}.

\subsection{The MOCL3 Runtime System}

When we implement the OpenCL runtime system on MT-3000, the key is to implement OpenCL APIs.
The OpenCL programming interfaces are mainly used to manage the interaction between the host and the accelerator,
including creating a context environment, managing the program object and compilation at the acceleration zone,
managing the buffers and data movements between the host zone and the acceleration zone,
starting the kernel program at the acceleration zone, etc.
Our OpenCL runtime system on MT-3000 is built on the heterogeneous driver and \htSysName (Figure~\ref{fig4:toolchain}).

\begin{figure}[!t]
\centering
\includegraphics[width=0.98\linewidth]{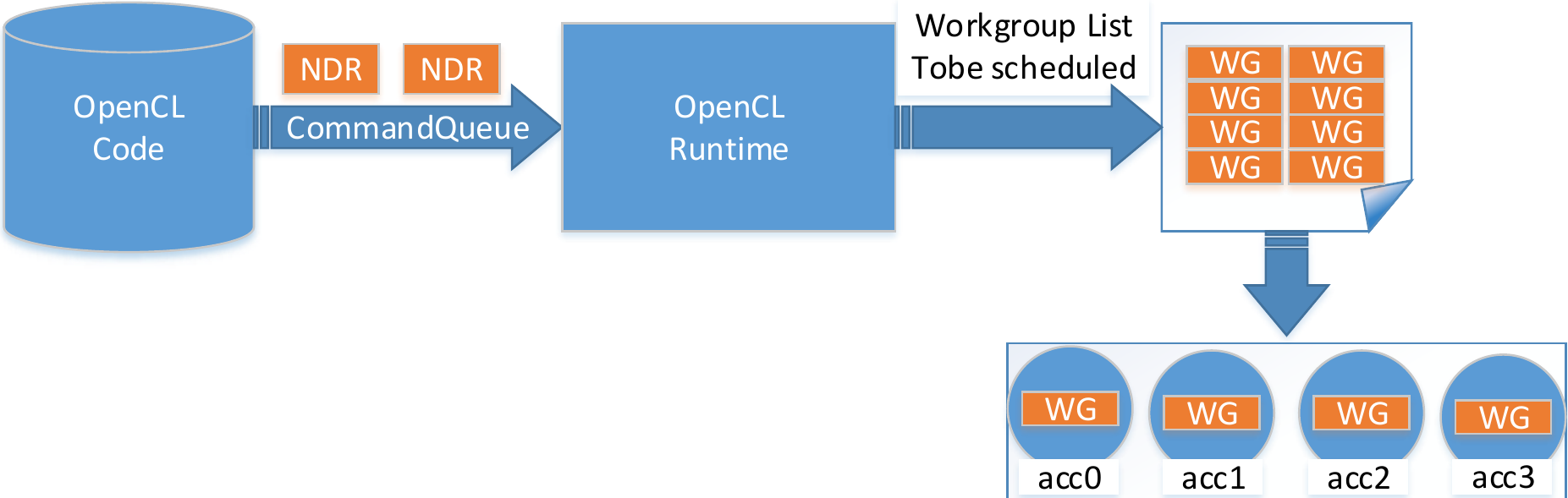}
\caption{The MOCL3 runtime system.}
\label{fig:ocl:runtime}
\end{figure}

According to the aforementioned OpenCL kernel compilation process,
a large number of concurrent tasks (i.e., workgroup functions) are defined by the index space.
After the kernel program is started at the accelerated zone,
the OpenCL runtime system needs to dispatch tasks during runtime. 
The process of executing an OpenCL program during runtime is shown in Figure~\ref{fig:ocl:runtime}.
Taking the OpenCL application as the input, the kernel (and the corresponding \emph{NDRange}) is submitted to the computing device
for execution through the OpenCL task queue, which is managed by the runtime system;
On the device side, a \emph{work group} is used as the basic unit to assign tasks to available acceleration array cores.
Considering that OpenCL program is of data parallel,
we use a static task mapping strategy.

\section{Results} \label{sec:results}
This section evaluates how \htSysName and \clSysName performs on MT-3000, and compares the design tradeoffs when implementing the programming interfaces for MT-3000.

\subsection{Conformance Test}
Since \htSysName is a new heterogeneous programming interface specifically targeting MT-3000,
we have developed a large number of internal test cases, a.k.a. the \texttt{ht-bench} suite,
which aims to cover all the APIs of \htSysName. Up to now, our \htSysName library can successfully verify all the test cases.

As for \clSysName, we use the test cases built in the POCL repository~\footnote{POCL, http://portablecl.org/}.
Our results demonstrate that \clSysName can pass all the 125 test cases covering the
ones of kernel compilation, runtime, work-groups, and regression.
This shows that \clSysName respects the OpenCL programming specification.

\subsection{Design Tradeoffs}
\subsubsection{Memory and Code Layout}
\begin{figure}[!t]
\subfigure[Memory layout.]{\includegraphics[width=0.8\linewidth]{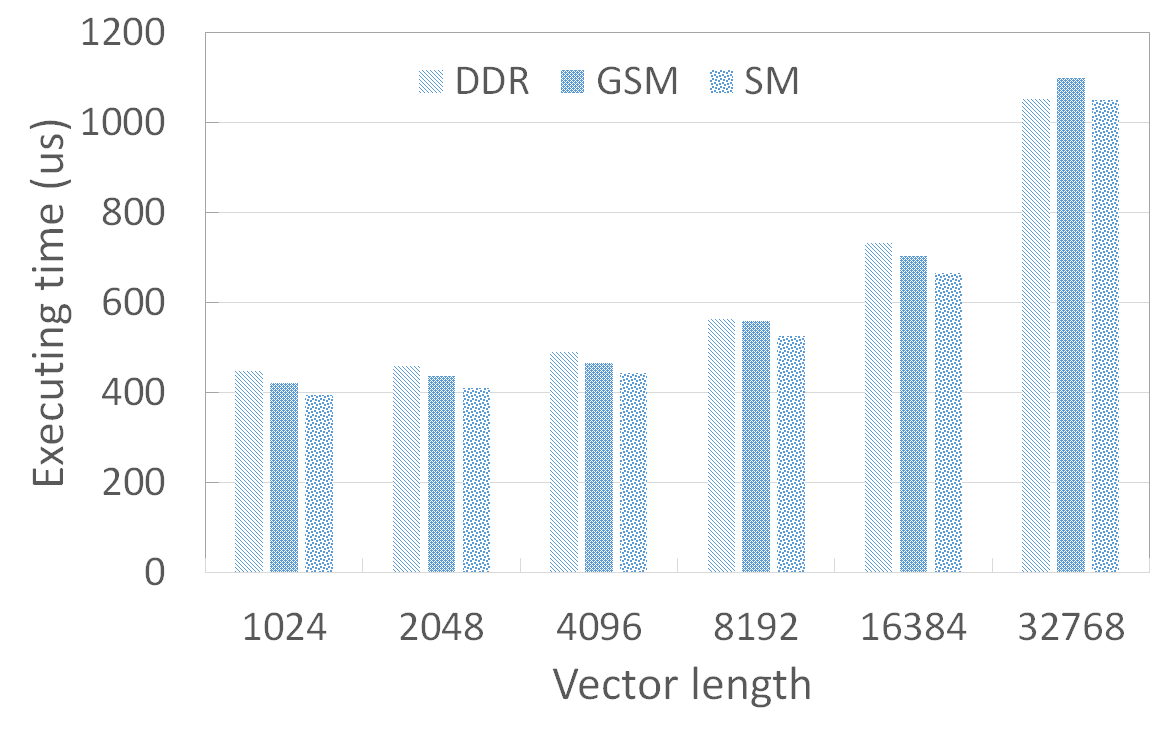} \label{fig:hthreads_sl}}
\subfigure[Code layout]{\includegraphics[width=0.8\linewidth]{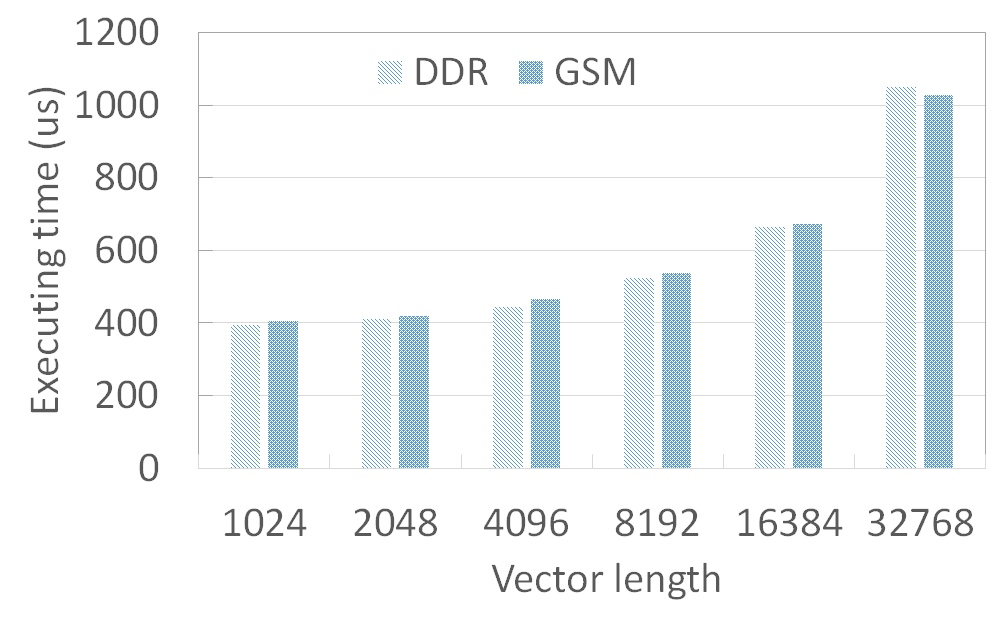} \label{fig:hthreads_cl}}
\caption{The performance comparison of different design choices: (a) The performance of placing stack on SM over DDR and GSM, (b) Performance comparison when placing code on DDR or GSM.}
\end{figure}

\cparagraph{Memory layout}
The placement of stack has a significant impact on performance.
As for MT-3000, the stack can be placed on either DDR, or GSM, or SM.
We have compared their performance with a micro-benchmark (i.e., a variant of vector addition) in \htSysName.
Figure~\ref{fig:hthreads_sl} shows the execution time of the three policies, i.e., placing stack on SM, GSM or DDR.
We see that placing the stack on SM yields the best performance, with an average performance improvement of 14\% over DDR.
Given that the micro-benchmark is simple,
we believe that this placing policy can achieve a larger performance speedup for the large real-life codes.

On the other hand, using the SM buffer is often beneficial for applications' performance by exploiting the data locality.
Thus we should use this on-chip buffer carefully, i.e., using it as stack or data cache.
To this end, we provide programmers with an environment 
variable \texttt{HT\_MEM\_LAYOUT\_POS=\{0, 0.2, 0.4, 0.6, 0.8, 1.0\}} in the production environment.
When it equals \texttt{0.2}, we use 20\% of SM per core as the stack space, and programmers can use the remaining 80\% space.
When \texttt{HT\_MEM\_LAYOUT\_POS=1.0}, we choose to use 100\% of SM per core as the stack space, and programmers have no access to this buffer.
In this way, programmers can fully utilize the SM buffers according to their applications.

\cparagraph{Code layout} We evaluate the performance impact of where the code segments are stored.
In MT-3000, we can map the code segments on either DDR or GSM.
Figure~\ref{fig:hthreads_cl} shows the performance of placing code on GSM over DDR.
We see that the performance benefit is within 5\%.
This is due to the fact that the acceleration core of MT-3000 has an instruction cache,
and the location of code segments has little impact on the overall performance.
Note that the GSM buffer can also be used as the on-chip data buffer to exploit data locality.
Therefore, we provide programmers with an environment variable \texttt{HT\_CODE\_LAYOUT\_POS=\{0, 1\}}.
When \texttt{HT\_CODE\_LAYOUT\_POS=0}, the code segments of \htSysName kernels are mapped onto the DDR space. Otherwise,
they are mapped to the GSM space.


\subsubsection{Launching Overhead} \label{sec:overhead}
One of the design goals of \htSysName is to minimize its management overhead, and 
hereby we evaluate the kernel launching overhead.
There are two ways to launch a kernel: one is to fire the acceleration cores directly, and the other is to use interrupt to wake up them.
Accordingly, there are two ways to check the kernel completion: one is to use query flags, and the other is to use interrupts.
In \texttt{libmt}, users can only start the kernel by firing the acceleration cores one by one, and then query the flags of each core.
In \htSysName, users can start the kernel by firing the acceleration cores with the \texttt{hthread\_group\_create} API, or
start it by interrupt with the \texttt{hthread\_group\_exec} API.
Checking the kernel completion in \htSysName is implemented by interrupts.

\begin{figure}[!t]
\centering
\includegraphics[width=0.8\linewidth]{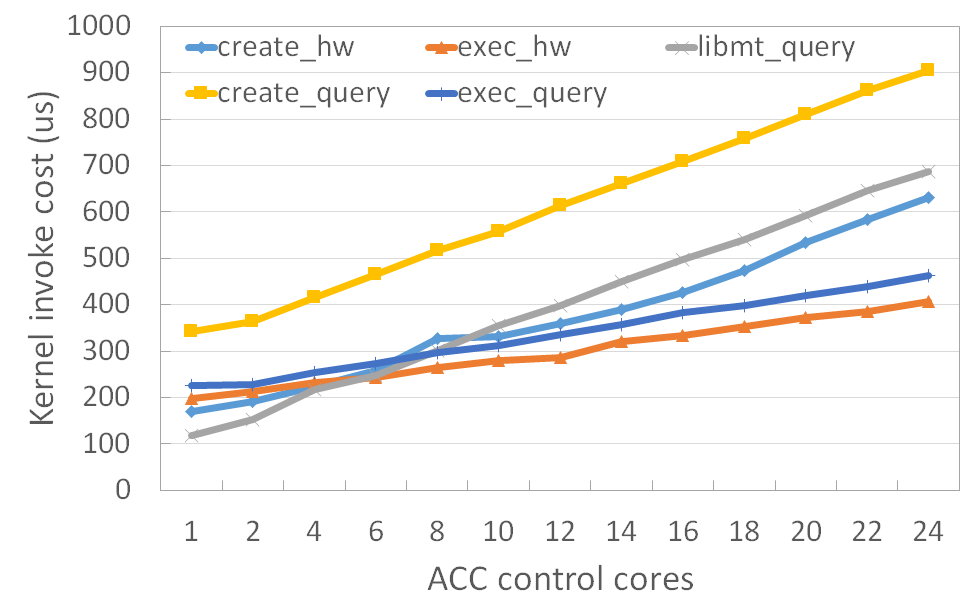}
\caption{The invoke cost of using hardware interrupt over querying.}
\label{fig:hthreads_ic}
\end{figure}

Since interrupts can be implemented either by hardware or simulated by querying flags,
we have evaluated the performance of both approaches.
Figure~\ref{fig:hthreads_ic} shows the 
launching overheads of using hardware interrupt over the querying approach.
We see that the overhead can be reduced by upto 50\% by starting kernel with interrupt compared with firing cores one by one.
The interrupt mode can be only supported by our OS-like runtime.
Using query flags to simulate interrupts increases overhead by 14\% compared to using hardware interrupts.
Compared to \texttt{libmt}, we can lower the launching overhead of using 24 cores from 700 us to 400 us.
When porting applications to Matrix-3000, programmers need to make sure that the kernel exetion time is much larger 
than the launching overhead, otherwise they can not obtain any performance benefit.
And they should fire the ACC zone once with the \texttt{hthread\_group\_create} API, 
and use the \texttt{hthread\_group\_exec} API to launch their kernel to minimize the launching overhead.


\subsection{Performance Optimization}
General matrix multiplication (GEMM) is a fundamental building block for high-performance computing (HPC) applications -
from traditional scientific simulations to emerging deep learning workloads. 
GEMM a matrix-multiply-accumulate operation, defined as $\mathbf{C}=\alpha \mathbf{A}
\cdot \mathbf{B} +\beta \mathbf{C}$, where $\mathbf{A}$ and $\mathbf{B}$ are matrix inputs, $\alpha$ and $\beta$ are scalar inputs, and
$\mathbf{C}$ is a pre-existing matrix which is overwritten by the output. 
Here matrix $\mathbf{A}$ is denoted as a \(M\times K\) matrix with $M$ rows and $K$ columns, matrix $\mathbf{B}$ is sized of $K
\times N$, and $\mathbf{C}$ is sized of $M \times N$.  
As a case study, we have implemented matrix multiplications in \htSysName and \clSysName. 

\subsubsection{Optimizing GEMM with \htSysName}
\begin{figure}[!t]
\subfigure[w/ \htSysName.]{\includegraphics[width=0.96\linewidth]{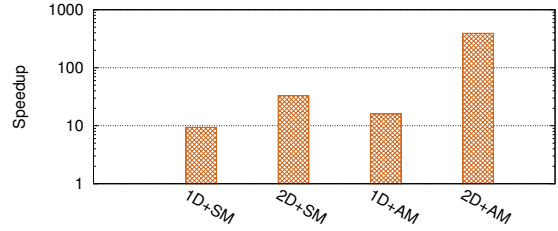} \label{fig:ht_mm}}
\subfigure[w/ \clSysName.]{\includegraphics[width=0.96\linewidth]{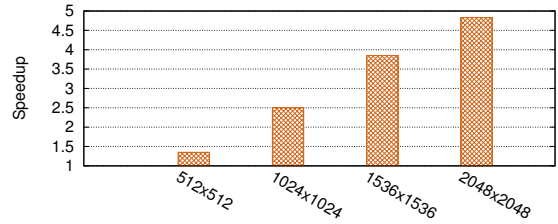} \label{fig:mocl3_lm}}
\caption{Speedup over baseline implementations for matrix multiplications: (a) with various optimizations in \htSysName,
(b) with local memory in \clSysName.}
\end{figure}

As a case study, we have implemented and optimized matrix multiplications with the \htSysName APIs.
The performance speedups of using various optimizations are shown in Figure~\ref{fig:ht_mm}.
We take the native parallel implementation with \htSysName as the baseline.
Based on this, we have mainly performed loop tiling (1D or 2D) and used the on-chip memory (SM or AM).
1D represents tiling the loop on one dimension, whereas 2D represents tiling the loop on two dimensions. 
To achieve data locality, we stage the tiled data on SM or AM.
The usage of on-chip buffers is achieved by exploiting the \htSysName DMA APIs to move data between the off-chip memory and the on-chip memory.
On Matrix-3000, we see that the achieved speedups of the four optimizations 
are 9.4$\times$, 33.04$\times$, 15.98$\times$, and 392.29$\times$, respectively.

\subsubsection{Optimizing GEMM with \clSysName}

We have also implemented matrix multiplications with \clSysName. 
Figure~\ref{fig:mocl3_lm} shows the performance improvement of using local memory over 
without local memory on MT-3000 for various inputs ($M=N=K=\{512, 1024, 1536, 2048\}$).
We see that by mapping the local memory to the on-chip memory of MT-3000, \clSysName can run matrix multiplications around 2.8$\times$ faster on average.
By moving data into the on-chip memories of MT-3000, we can improve the memory bandwidth by exploiting the fast on-chip buffers and reusing the data elements.
Note that, we use the \texttt{async\_work\_group\_copy} \emph{builtins} to move data from global memories to local memories.

%

To summarize, we have used matrix multiplications as a case study to demonstrate the performance of \htSysName and \clSysName. 
We find that the programming model developed for Matrix-3000 performs well in terms of both \textit{performance} and \textit{programmability}.

\section{Related Work}
Parallel programming model acts as the bridge between programmers and
parallel architectures.
To use the shared memory parallelism on multi-core CPUs,
parallel programming models are often implemented on threading mechanisms such as the POSIX threads~\cite{DBLP:conf/usenix/Alfieri94}.
For programming heterogeneous many-cores, 
Fang \etal have summarized
the family of parallel programming models for heterogeneous many-core architectures~\cite{DBLP:journals/ccfthpc/FangHTW20}.
Based on the performance-programmability tradeoff,
the programming models/languages are categorized into
\textit{low-level programming models} and \textit{high-level programming models}.
The expected application \textit{performance} increases from high-level programming models to low-level programming models,
whereas the \textit{programmability} decreases.

The low-level programming models are closest to the many-core architectures, and
expose the most hardware details to programmers through data structures and/or APIs.
These models are typically bound to specific hardware architectures, and are also
known as \textit{native programming models}.
The representative models are libSPE for STI Cell/B.E.~\cite{book:cellbe}, CUDA for NVIDIA GPUs\footnote{https://developer.nvidia.com/cuda-downloads}.
In contrast, the high-level programming models raise the languages' abstraction level,
and hide more architecture details than the low-level models.
Thus, the high-level models often enable better programmability.
The representative models are SYCL\footnote{https://www.khronos.org/sycl/}, Kokkos~\cite{DBLP:journals/tpds/TrottLACDEGHHIL22}, 
OpenACC\footnote{https://www.openacc.org/}, and PyTorch\footnote{https://pytorch.org/}.

To achieve high performance, high programmability, and high portability, 
we argue that a holistic solution of programming systems is required for future heterogeneous many-cores. 
This article shares our experiences on the development of programming systems for our home-grown heterogeneous processor.
At the low level, we present a close-to-metal programming interface (\texttt{hthreads}) to tap the hardware potentials. 
At the high level, we present a customized implementation of the standards programming interface (\texttt{MOCL3}). 
This holistic solution aims to achieve a balance between performance, programmability, and portability.

\section{CONCLUSION}
We have presented the design and implementation of the programming and compiler tools for the Matrix-3000 accelerator. 
The complex memory hierarchy and processor core organization of Matrix-3000 require software to make
effective use of the microarchitecture design to realize the hardware performance. 
We share our experience on how a low-level threading-based programming interface 
can be developed to support the high-level OpenCL programming standard.  
We hope the experience shared in this article can support the design and implementation of systems software
for future specialized computing hardware.

\balance
\bibliographystyle{fitee}
\bibliography{ref}

\end{document}